\documentclass[a4paper,11pt]{JHEP3}
\usepackage{amsmath}
\preprint{SISSA 35/2006/EP}
\numberwithin{equation}{section}
\allowdisplaybreaks
\title{\large The world-sheet corrections to dyons in the Heterotic theory}
\author{{Qasem Exirifard}
\\
Institute for studies in Theoretical Physics and Mathematics (IPM),\\  P.O.Box 19395-5531, Tehran, Iran.\\
Email: \email{exir@theory.ipm.ac.ir}
}

\abstract{All the linear $\alpha'$ corrections, however excluding the gravitational Chern-Simons correction, are studied in the toroidally compactified critical Heterotic string theory.  These corrections are computed to the entropy for a BPS static spherical four dimensional dyonic black hole which represents a wrapped fundamental string  carrying arbitrary  winding and  momentum charges along one cycle in the presence of KK-monopole and H-monopole charges associated to another cycle. It is verified that after the inclusion of the gravitational Chern-Simons corrections \cite{Bindusar}, all the linear $\alpha'$ corrections to the entropy for the supersymmetric dyon can be reproduced by the inclusion of only the Gauss-Bonnet Lagrangian to the supergravity approximation of the induced Lagrangian.}

\begin{document}
\bibliographystyle{JHEP}
\section{Introduction}
Dyonic black holes are black holes which carry electric  and magnetic charges of some gauge fields. Some of the dyonic black holes can be realised as the solutions to the supergravity approximation to the critical Heterotic string theory compactified on $T^6$. The proposal for the exact degeneracy of microsates of dyons in the toroidally compactified critical Heterotic string theory 
\cite{Dijkgraaf:1996it}-\cite{Shih:2005he} 
with a negligible string coupling constat recently has been proved in the ref.\,\cite{David:2006yn}. The logarithm of the degeneracy of dyons defines the statistical entropy.

At the supergravity approximation the Hawking-Bekenstein entropy is in agreement with the large-charge-limit of the statistical entropy. The dominant string corrections to these dyons, the $\alpha'$ corrections, should be in agreement  with the large charge expansion series of the statistical entropy. Ref.\,
\cite{Behrndt:1998eq}-\cite{LopesCardoso:2000fp}
 observed that upon the inclusion of the square of the Riemann tensor and a supersymmetric completion of that, the modified Hawking-Bekenstein entropy \cite{wald,wald2} is in agreement with the statistical entropy. Ref.\,\cite{int4} has shown that the inclusion of the Gauss-Bonnet action gives the same corrections in modified Hawking-Bekenstein entropy as those given by the inclusion of the supersymmetric version of the square of the Riemann tensor.

The Gauss-Bonnet Lagrangian or the supersymmetric version of the square of the Riemann tensor are not all the linear $\alpha'$ corrections to the dyonic black holes.  A priori it is not clear that why other linear $\alpha'$ should not contribute to the entropy. Ref.\,\cite{Kraus1,Kraus2} has proved certain non-renormalisation theorems establishing that for a certain family of supesymmetric black holes in the Heterotic string theories the results of the ref.\,
\cite{Behrndt:1998eq}-\cite{LopesCardoso:2000fp} are in fact exact.

This work aims to  verify the results of the ref.\,\cite{Kraus1,Kraus2} versus a manifest calculation which takes into account all the four derivative corrections for an extremal supersymmetric dyonic black hole. In the first preprint \cite{Qasem} I left out the coupling of the gravitational Chern-Simons terms to the three form field strength. The authors of the ref.\,\cite{Bindusar} realised this left out and  they further computed the gravitational Chern-Simons corrections to the entropy, based on the frame work which they had developed earlier \cite{Bindusar2}. Here I shall calculate all the four derivative corrections, however excluding the  gravitational Chern-Simons terms, to the entropy. I verify that, as ref.\,\cite{Bindusar} concludes, after the inclusion of the gravitational Chern-Simons terms ``the resulting entropy  agrees perfectly with the results of earlier analysis, in accordance with the non-renormalisation theorems of the ref.\,\cite{Kraus1,Kraus2}''. The note is organised in the following way:

In the second section we consider the Low Energy Effective Action of the Heterotic string theory. We study a KK-compactification of the Heterotic string theory on $T^6$ which is relevant for a BPS static spherical four dimensional dyonic black hole representing a wrapped fundamental string  carrying arbitrary  winding and  momentum charges along one cycle in the presence of the KK-monopole and the H-monopole charges associated to another cycle. This solution is constructed in the ref.\,\cite{Cvetic:1995uj}.

The third section  presents all the linear $\alpha'$ corrections in the critical Heterotic string theory for the Bosonic (NS) backgrounds derived from the string amplitude considerations on a sphere \cite{stringamplitude1,stringamplitude2}.  The compactification process of the second section will be employed to account for all the linear $\alpha'$ corrections in the induced action. The $\alpha'$ corrections will be studied as perturbations outside the horizon. A boundary condition, hereby referred to as the H-boundary conditions, will be imposed on the linear $\alpha'$ corrections to insure that the $\alpha'$ corrections remain bounded on and outside the horizon. It will be noticed that the H-boundary conditions does not necessarily retain the fall off of the fields at the asymptotic infinity. Therefore the charges may receive corrections when the H-boundary conditions are imposed. The charges of the dyonic black hole retain their values  because there exists a scheme in which the dyonic black hole does not receive any $\alpha'$ corrections \cite{Cvetic:1995yq, Horowitz:1994rf}.

In the fourth section, having excluded the gravitational Chern-Simons terms, we shall calculate the induced action near the horizon configuration. We employ the attractor mechanism \cite{Ferrara:1995ih,Strominger:1996kf}  and the entropy formalism \cite{int4,int5} to calculate the modified Hawking-Bekenstein entropy. We will see that when we add the gravitational Chern-Simons corrections \cite{Bindusar} to the results then the linear $\alpha'$ corrections to the thermodynamical entropy is both in perfect agreement with the statistical entropy, and  in accordance  with the non-renormalisation theorems of the ref.\,\cite{Kraus1,Kraus2} .

\section{The compactification of the LEEA}
Let a ten-dimensional Riemannian manifold ${\boldsymbol{M}}_{10}$ homeomorphic to $M_4\times T^{6}$ be considered whose metric has six killing vectors in $T^6$ and $\boldsymbol{M}_{10}$ admits an asymptotically flat region. Let  $\boldsymbol{M}_{10}$ be endowed with a two-form and a scalar fields.  We represent the coordinate patch that covers the asymptotic flat region of $\boldsymbol{M}_{10}$ by $\boldsymbol{x}^{i}$ =$(t$, $x^1$, $x^2$, $x^3$, $y_1$, $y_2$, $z_1$, $\cdots$, $z_4)$ where $x^{\mu}=(t,\cdots,x^3)$ and $(y^m,z^m)=(y_1,y_2,z_1,\cdots,z_4)$ are coordinates respectively on $M_4$ and $T^6$ in such a way that $(dy_1,\cdots,dz_4)$ are the killing vectors. We refer to $x_\mu$ and ($y^m$,$z^m$) respectively as the four dimensions and the compactified space. The string theory can be  studied perturbatively at the vicinity of the asymptotic region of  $\boldsymbol{M}_{10}$ which is covered by $\boldsymbol{x}^i$. We look at this vicinity as a background of the metric, the NS-two form and the dilaton for the propagation of the critical Heterotic string theory,
\begin{eqnarray}
{\boldsymbol{ds^2}}&=&\sum_{\scriptscriptstyle{\mu,\nu=1}}^{4} \boldsymbol{g}_{\mu\nu} (x) dx^{\mu} dx^{\nu} +  \sum_{\scriptscriptstyle{m=1}}^{2}\{2\boldsymbol{g}_{y_m\mu}(x)\,dy_m\, dx^{\mu}+\boldsymbol{g}_{y_my_m}(x) dy_m^2\} +\sum_{\scriptscriptstyle{m=1}}^{4} dz_{m}^2\,,\nonumber\\
{\boldsymbol{B}}&=&\boldsymbol{B}_{\mu\nu}(x)dx^{\mu}\wedge dx^{\nu}+\boldsymbol{B}_{y_1\mu}(x) dx^{\mu}\wedge dy_1+\boldsymbol{B}_{y_2\mu}(x) dx^{\mu}\wedge dy_2\,,\\
{\boldsymbol{\phi}}&=&\boldsymbol{\phi}(x)\,.
\end{eqnarray}
We use the bold symbols to represent the fields in the ten dimensions. We rewrite the metric in the following form
\begin{equation}\label{themetric}
\boldsymbol{d}\boldsymbol{s}^2 = g_{\mu\nu}(x)\, dx^\mu dx^\nu + T_1(x)^2 (dy_1+2  A^1_{\mu}(x)dx^{\mu})^2+ T_2(x)^2(dy_2+2 A_\mu^2(x)dx^\mu)^2+ dz_m^2\,, 
\end{equation}
where $g_{\mu\nu}$, $T_1$, $T_2$, $A^1_\mu$ and $A^2_\mu$ are employed to re-express the components of the ten dimensional metric in such a way that we shall see is more convenient. The metric (\ref{themetric}) is invariant under the following transformations
\begin{equation}
\left\{\begin{array}{l}
y_1 \to y_1 - 2 \Lambda^1(x)\\
A^1_\mu(x)\to A^1_\mu(x) + \partial_\mu \Lambda^1(x)
\end{array}
\hspace{1cm},
\right\{\begin{array}{l}
y_2 \to y_2 - 2 \Lambda^2(x)\\
A^2_\mu(x)\to A^2_\mu(x) + \partial_\mu \Lambda^2(x)
\end{array}
\end{equation}
where $\Lambda^1(x)$ and $\Lambda^2(x)$ are arbitrary scalars. These symmetries can be thought as remnants of the ten-dimensional diffeomorphism within the compactification process. We interpret $A^1_\mu(x)$ and $A^2_\mu(x)$ as two distinct $U(1)$ gauge connections in the four dimensions because they are vectors and they have $U(1)$ symmetries associated to them.

We rewrite the NS two-form in the following way
\begin{equation}\label{theB}
\boldsymbol{B} = B_{\mu\nu} dx^\mu\wedge dx^\nu 
+ 2 A^3_\mu dx^\mu \wedge (dy_1+ 2 A^1_\nu dx^\nu) 
+ 2 A^4_\mu dx^\mu \wedge (dy_2+ 2 A^2_\nu dx^\nu)\,,
\end{equation}
where $B_{\mu\nu}$,  $A^3_\mu$ and $A^4_\mu$ are used to re-express the components of the ten dimensional NS two-form in such a form to guaranty the  convenience in forthcoming computations. Note that the $U(1)$ transformations associated to $A^1_\mu(x)$ and $A^2_\mu(x)$ leave  intact $B_{\mu\nu}(x)$,  $A^3(x)$ and $A^4(x)$ because $dy_i+ 2 A^i_\nu dx^\nu$  is gauge invariant. The sigma model for the background we are considering is invariant under altering $\boldsymbol{B}$ by any exact two-form, i.e.
\begin{equation}
 \boldsymbol{B} \to \boldsymbol{B}+ \boldsymbol{d}\boldsymbol{\Lambda}\,,
\end{equation}
and thus the low energy action is invariant under the symmetry transformation given by $\boldsymbol{B} \to \boldsymbol{B}+ \boldsymbol{d}\boldsymbol{\Lambda}\,$.  A subset of these transformations is given by 
\begin{displaymath}
\boldsymbol{\Lambda}=\Lambda_\mu(x) dx^\mu - \Lambda^3(x) dy_1 - \Lambda^4(x) dy_2\,,
\end{displaymath}
and this subset implies that the LEEA is invariant under
\begin{eqnarray}
B  &\to& B+ d \Lambda\,,\\
A^3_\mu(x)&\to& A^3_\mu(x) + \partial_\mu \Lambda^3(x)\,,\\ 
A^4_\mu(x)&\to& A^4_\mu(x) + \partial_\mu \Lambda^4(x)\,.
\end{eqnarray}
We see that independent $U(1)$ symmetries are associated to $A^3_\mu(x)$ and $A^4_\mu(x)$. These $U(1)$ symmetries are remnants of the gauge symmetries in the ten-dimensions. We interpret $A^3_\mu(x)$ and $A^4_\mu(x)$ as two new distinct gauge connections in the four dimensions.

Because of the symmetries of the metric we can choose a sufficiently large volume for any non-trivial cycle\footnote{A cycle which does not shrink to a point under any given homeomorphism.} in the compactified space within the patch of $\boldsymbol{x}^i$. Therefore we consistently ignore the world-sheet or the target space instantons corrections to the LEEA. The rest of the string corrections to LEEA respects the ten dimensional diffeomorphism symmetry group. Therefore the ten-dimensional low energy effective action can be written by
\begin{equation}\label{10Daction}
\boldsymbol{S}[\boldsymbol{B},\boldsymbol{g},\boldsymbol{\phi}]\, =\,\frac{1}{32\,\pi}\, \int\,d^{10}\boldsymbol{x}\sqrt{-\det \boldsymbol{g}}\, e^{-2\boldsymbol{\phi}}\, \boldsymbol{L}(\boldsymbol{B},\boldsymbol{g},\boldsymbol{\phi})\,, 
\end{equation}
where $\boldsymbol{S}[\boldsymbol{B},\boldsymbol{g},\boldsymbol{\phi}]$ includes all the perturbative string corrections and it is invariant under the ten-dimensional diffeomorphism and $\boldsymbol{B} \to \boldsymbol{B}+ \boldsymbol{d}\boldsymbol{\Lambda}$. 

Due to the presence of the gravitational Chern-Simons terms the Lagrangian density - $\boldsymbol{L}(\boldsymbol{B},\boldsymbol{g},\boldsymbol{\phi})$-  is not generically invariant under both the ten dimensional diffeomorphism and  gauge symmetry groups. Here we are going to calculate all the linear $\alpha'$ corrections but the gravitational Chern-Simons corrections, therefore we consider the perturbative string corrections which does not include the  gravitational Chern-Simons terms. From this time on we shall consider only this subset of the corrections. Within this subset $\boldsymbol{L}(\boldsymbol{B},\boldsymbol{g},\boldsymbol{\phi})$ can be written in a covariant form in terms of the ten dimensional Riemann tensor, the exterior derivative of $\boldsymbol{B}$, the dilaton and their covariant derivatives respect to the ten dimensional metric,
\begin{eqnarray}
\boldsymbol{L}(\boldsymbol{B},\boldsymbol{g},\boldsymbol{\phi})&=& \boldsymbol{L}(\boldsymbol{R}_{ijkl},\boldsymbol{d}\boldsymbol{B},\boldsymbol{\nabla})\\
\boldsymbol{S}[\boldsymbol{B},\boldsymbol{g},\boldsymbol{\phi}]& =&\frac{1}{32\,\pi}\, \int\,d^{10}\boldsymbol{x}\sqrt{-\det \boldsymbol{g}}\, e^{-2\boldsymbol{\phi}}\, \boldsymbol{L}(\boldsymbol{B},\boldsymbol{g},\boldsymbol{\phi})\,, \label{10Daction1}
\end{eqnarray}
To find the pullback of the action to four dimensions we insert eq.\,(\ref{themetric}) and (\ref{theB}) into the action (\ref{10Daction1}) and we obtain
\begin{equation}\label{4daction}
S\, = \,\frac{1}{32\,\pi}\,\int d^4x  \, \sqrt{-\det g}\, e^{-2\phi}\, L(g_{\mu\nu},B_{\mu\nu}, A^1_\mu,\cdots,A^4_\mu,T_1,T_2,\phi)\,,
\end{equation}
where the integration on the compactified space is understood and we have defined 
\begin{equation}\label{thedilaton}
2 \phi = 2 \boldsymbol{\phi} - \ln T_1 - \ln T_2 - \ln V\,,
\end{equation}
and $V$ is the volume of the compactified space. We refer to the eq. (\ref{4daction}) as the induced action. The induced action and the induced lagrangian inherit the remnants symmetries of the ten dimensional action. Thus they are invariant under the four dimensional diffeomorphism group, four $U(1)$ symmetries associated to $A^1_\mu,\cdots,A^4_\mu$ and $B\rightarrow B+d\Lambda$. This means that the induced Lagrangian, for the considered subset of the corrections, is expressible in a covariant form in terms of the Riemann tensor constructed from $g_{\mu\nu}$, the exterior derivatives of the form fields, $T_1$, $T_2$, $\phi$ and their covariant derivatives respect to $g_{\mu\nu}$, 
\begin{eqnarray}\label{inducedlagrangian}
{L}&=&{L}(R_{\mu\nu\lambda\eta}, dB, dA^1, \cdots, dA^4, T_1, T_2, \phi, g_{\mu\nu},\nabla_\mu)\,\blacksquare
\end{eqnarray}

\section{Dyonic black holes and the $\alpha'$ corrections}
In the supergravity approximation to the critical Heterotic string theory the action for backgrounds of the metric ($\boldsymbol{g}$), the NS two-form ($\boldsymbol{B}$) and the dilaton ($\boldsymbol{\phi}$ ) reads
\begin{eqnarray}\label{treelevelaction}
\boldsymbol{S}^{(0)}&=&\frac{1}{32 \pi }\int d^{10}\boldsymbol{x}\, \sqrt{-\det\boldsymbol{g}}\,e^{-2\boldsymbol{\phi}}\,({\boldsymbol{R}_{\scriptscriptstyle{\text{Ricci}}}} + 4 |\boldsymbol{\nabla} \boldsymbol{\phi}|^2 - \frac{1}{12} \boldsymbol{H}_{ijk}\boldsymbol{H}^{ijk})\,,
\end{eqnarray}
where
\begin{eqnarray}\label{defineH}
\boldsymbol{H}_{ijk}&=& \boldsymbol{d}\boldsymbol{B}\,=\, {3}\, \boldsymbol{B}_{[ij,k]}\,.
\end{eqnarray}
The linear $\alpha'$ corrections to this action, derived from three \cite{5Gross,5Nishino,5Cai} and four \cite{stringamplitude2,5Green, 5Schwarz} vertex operator insertions \cite{vertex}  in the string amplitude considerations on sphere, are\footnote{Note that ref.\,\cite{stringamplitude2} computed the linear $\alpha'$ corrections for the Heterotic and the Bosonic string theory. Eq. (\ref{LMT}) is  eq.\,(3.24) of the ref.\,\cite{stringamplitude2} for $\lambda_0=\frac{1}{8}$.}
\begin{eqnarray}\label{LMT}\nonumber
\boldsymbol{S}^{(1)}&=&\frac{1}{32 \pi }\int d^{10}\boldsymbol{x} \sqrt{-\det\boldsymbol{g}}~e^{-2\boldsymbol{\phi}}\,\frac{\alpha'}{8}\,\left( {\boldsymbol{R}_{klmn}\boldsymbol{R}^{klmn}} 
-\frac{1}{2}\boldsymbol{R}_{klmn}\boldsymbol{H}_{p}^{~kl}\boldsymbol{H}^{pmn} \,\right.\\
&&\left.\hspace{2cm}-\frac{1}{8}\boldsymbol{H}_{k}^{~mn}\boldsymbol{H}_{lmn}\boldsymbol{H}^{kpq}\boldsymbol{H}^{l}_{~pq}
+\frac{1}{24}\boldsymbol{H}_{klm}\boldsymbol{H}^{k}_{~pq}\boldsymbol{H}_{r}^{~lp}\boldsymbol{H}^{rmq}\right)\,,
\end{eqnarray}
plus the Lorentz Chern-Simons term that could be reproduced  by substituting $\boldsymbol{H}$  in the eq.\,(\ref{treelevelaction}) with 
\begin{eqnarray}\label{HH}
\boldsymbol{H}^{\star}&=& \boldsymbol{d}\boldsymbol{B}+ \frac{\alpha'}{4}\,\boldsymbol{\omega}_{3L}(\boldsymbol{\Omega})\,,
\end{eqnarray}
where $\boldsymbol{\omega}_{3L}(\boldsymbol{\Omega})$ encodes the gravitational Chern-Simon modification to $\boldsymbol{d}\boldsymbol{B}$. It is the insertion of three vertex functions that produces the Lorentz gravitational Chern-Simons term. Due to  the consistency of the Heterotic string around the flat space-time the eq.\,(\ref{LMT}) being derived from the string amplitude considerations should be in agreement with the related worldsheet-loop calculations of the corresponding sigma model. The corrections given by the eq.\,(\ref{LMT}) are all the four derivative corrections to the action. Thus the eq.\,(\ref{LMT}) should also be compatible with the anomaly cancellation mechanism \cite{DBranes} for the backgrounds of the vanishing $\text{Spin}(32)/Z_2$ or $E8 \times E8$ connection. 

The contribution of the gravitation Chern-Simons corrections to the entropy for the dyon we are intersted in has been computed in the ref.\,\cite{Bindusar}. In this note we study all the linear $\alpha'$ corrections but the gravitational Chern-Simons correction to the entropy.   So  we do not include the gravitational Chern-Simons modification to $\boldsymbol{d}\boldsymbol{B}$ in the forthcoming computations and we shall use the eq.\,(\ref{defineH}) rather than the eq.\,(\ref{HH}). At the end we add the gravitation Chern-Simons corrections to the entropy \cite{Bindusar}.

We compactify the ten dimensional background to the four dimensions in the way that we presented in the previous section. This means that we rewrite the ten dimensional background by eq. (\ref{themetric}), (\ref{theB}) and (\ref{thedilaton}) in terms of the four dimensional background fields; the metric $g_{\mu\nu}$, the two-form $B_{\mu\nu}$, the four $U(1)$ gauge connections $A^1 \cdots A^4$ and the three scalars $T_1$, $T_2$ and $\phi$. It is not a hard task to obtain the explicit form of the induced action at the level of supergravity approximation,
\begin{eqnarray}\nonumber
S^{(0)}&=&\frac{1}{32\mathit{\pi}} \int\, d^4x\, \sqrt{-\det g}\,e^{-2\phi} 
\left(R- \frac{|dB|^2}{12}+4 |\nabla \phi|^2 -  |\nabla \ln T_{1}|^2 - |{\nabla \ln T_{2}}|^2\right.\\
&&\hspace{3cm}\left.
-|T_{1}\,d A^{1}|^2 
-|T_{2}\,d A^{2}|^2
-|\frac{d A^{3}}{T_{1} }|^2
-|\frac{d A^{4}}{T_{2} }|^2\right)\,,
\end{eqnarray}
where $R$ denotes the Ricci scalar of $g_{\mu\nu}$ and integrations by parts are understood. We do not obtain the explicit form of the linear $\alpha'$ corrections to the induced action. We suffice to present the linear $\alpha'$ corrections to the induced action by
\begin{eqnarray}
S^{(1)}&=& \frac{1}{32\, \mathit{\pi} }\int d^4x \, \sqrt{-\det g}\,e^{-2\phi} \,L^{(1)} \,,
\end{eqnarray}
and we know that $L^{(1)}$ is a functional of the four dimensional Riemann tensor, the gauge field strengths and their covariant derivatives,
\begin{eqnarray}
L^{(1)} &=&L^{(1)}(R_{\mu\nu\lambda\eta}, dB, dA^1, \cdots, dA^4, T_1, T_2, \phi, g_{\mu\nu},\nabla_\mu)\,.
\end{eqnarray}
At the level of the four dimensional supergravity approximation we consider a static spherical dyonic black hole which carries electric charges of $A^1$ and $A^3$ and magnetic charges of $A^2$ and $A^4$ in an asymptotically flat space-time. The electric charges of $A^1$ and $A^3$ represent respectively the KK-momentum and winding numbers of a fundamental string wrapped around the cycle of $y_1$. The magnetic charges of $A^2$ and $A^4$ represent respectively the KK-monopole and the H-monopole charges associated to the cycle of $y_2$. The explicit forms of the fields for this dyonic black hole are presented in the ref.\,\cite{Cvetic:1995uj}. When none of the charges is zero then the dyonic black hole has a regular horizon with geometry of $AdS_2\times S^2$ outside which the string loop corrections can be ignored. At the supergravity approximation the $SO(6,22)\times SL(2,Z)$ duality transformations can be applied on the dyonic black hole to obtain a general dyonic black hole \cite{SenOdd}. Reminiscent to T-duality and $\alpha'$ corrections \cite{martin,exir2,T1}, we expect that the duality transformations themselves get modified by the $\alpha'$ corrections. In this note we do not use the duality transformations. We consider a dyonic black hole with large  momentum, winding, KK-monopole and H-monopole charges and we study the $\alpha'$ corrections as perturbations on and outside its horizon.

Since we do have so many fields in the four dimensions we use a compact notation. Let $\Psi_i$ represent all the fields of the dyonic black hole in a compact fashion\footnote{This collective notation is constructed in analogy with the compact notation used in the ref.\,\cite{Seif}.},
\begin{eqnarray}
 \Psi_i &\in& \left\{ g_{\mu\nu}(x), T_1(x), T_2(x),  A_\mu^{1}(x), \cdots, A_\mu^{4}(x), \phi (x)\right\}\,.
\end{eqnarray}
The action for this compact notation follows
\begin{eqnarray}
S[\Psi] &=& S^{0}[\Psi] + \alpha' S^{1}[\Psi] + O(\alpha'^2)\,, 
\end{eqnarray}
the equations of motion of which read 
\begin{eqnarray}\label{PsiEq}
0\,=\,\dfrac{\delta S[\Psi]}{\delta \Psi_i} &=& \dfrac{\delta S^0[\Psi]}{\delta \Psi_i} + \alpha' \dfrac{\delta S^1[\Psi]}{\delta \Psi_i} + O(\alpha'^2)\,,
\end{eqnarray}
where $\frac{\delta{~}}{\delta \Psi_i}$ stands for the functional derivative respect to $\Psi_i$. Let us write an $\alpha'$ expansion series for $\Psi_i$ and next solve the eq.\,(\ref{PsiEq}) perturbatively,
\begin{eqnarray}
 \Psi_i & = & \Psi_i^0 + \alpha' \Psi^1_i + O(\alpha'^2) \,.
\end{eqnarray}
Inserting this perturbative expansion in the eq.\,(\ref{PsiEq})  gives
\begin{eqnarray}
\dfrac{\delta S^0[\Psi]}{\delta \Psi_i} \big{|}_{_{\alpha'=0}}+
\alpha' (\dfrac{\delta^2 S^0[\Psi]}{\delta \Psi_i\delta\Psi_j}\big{|}_{_{\alpha'=0}}\Psi^1_j+\dfrac{\delta S^1[\Psi]}{\delta \Psi_i}\big{|}_{_{\alpha'=0}}) + O(\alpha'^2)\,=\,0\,,
\end{eqnarray}
which implies 
\begin{eqnarray}\label{SEQ}
\dfrac{\delta S^0[\Psi]}{\delta \Psi_i} \big{|}_{_{\alpha'=0}} &=& 0\,,\\\label{EQAlpha}
\dfrac{\delta^2 S^0[\Psi]}{\delta \Psi_i\delta\Psi_j}\big{|}_{_{\alpha'=0}}\Psi^1_j &=&-\dfrac{\delta S^1[\Psi]}{\delta \Psi_i}\big{|}_{_{\alpha'=0}} \,.
\end{eqnarray}
Note that the eq.\,(\ref{SEQ}) stands for the equations of motion at the supergravity approximation and the eq.\,(\ref{EQAlpha}) gives a set of non-homogeneous linear second order differential equations for $\{\Psi_i^1\}$ for any given solution at the supergravity approximation $\{\Psi_i^0\}$. 

Let us first study the solutions to the homogeneous equations corresponding to the eq.\,(\ref{EQAlpha}), 
\begin{eqnarray}\label{HomogeneousEquations}
\dfrac{\delta^2 S^0[\Psi]}{\delta \Psi_i\delta\Psi_j}\big{|}_{_{\alpha'=0}}\Psi^1_{j,B} &=&0\,. 
\end{eqnarray}
These are the equations describing the fluctuations around $\Psi=\Psi_0$ at the supergravity approximation. For the dyonic black hole, the equation for a static spherical fluctuation of the dilaton in the canonical frame is simplified to
\begin{eqnarray}\label{HODILEQ}
\partial_r ((r-r_{\scriptscriptstyle{H}})^2\partial \phi^{1}_B) &=&0\,,
\end{eqnarray}
where $r_{\scriptscriptstyle{H}}$ is the radius of the horizon and $\phi^{1}_B$ is the fluctuation of the dilaton and we have used the explicit form of the background fields presented in the ref.\,\cite{Cvetic:1995uj}. The general solution of the eq.\,(\ref{HODILEQ}) is diverging on the horizon,
\begin{eqnarray}
\phi^{1}_B (r) &=& \frac{c_1}{r-r_{\scriptscriptstyle{H}}}+c_2\,. 
\end{eqnarray}
The diverging mode of the dilaton fluctuations plays the role of the diverging source for the fluctuations of all other fields through their couplings to the dilaton. Thus the fluctuations of all other fields admit modes which  diverge on the horizon. Therefore we conclude that 
\begin{quotation}
\hspace{-0.7cm} {\bf Lemma 1:} The general solutions to the homogeneous equations (\ref{HomogeneousEquations})  diverge on the horizon.
\end{quotation}
The  diverging  solutions on the horizon should be excluded by the boundary conditions if we are going to have a smooth solution outside the horizon. Let the following boundary conditions be introduced for the solutions of the eq.\,(\ref{EQAlpha}) 
\begin{eqnarray}\label{Hboundary}
\left\{
\begin{array}{rcl}
\Psi_i^1(x)\big{|_{_{x=\infty}}}&=&0\,,\\ \\
\Psi_i^1(x)\big{|_{_{x  \text{~on the Horizon}}}}&<&\infty \,, 
\end{array}
\right.
\end{eqnarray}
We call the above boundary conditions  as the H-boundary conditions. The first condition of the H-boundary conditions sets the $\alpha'$ corrections to zero at infinity and its second condition excludes the diverging modes on the horizon. Depends on how we decide to represent the metric, some of the components of the metric may diverge on the horizon. For these components of the metric we  substitute the second condition of the H-boundary conditions by 
\begin{eqnarray}\label{3.21}
 \lim_{ x\rightarrow\Sigma_h}\,\dfrac{\Psi_i^1(x)}{\Psi_i^0(x)}\,<\,\infty\,,
\end{eqnarray}
where $\Sigma_h$ represents any point on the event horizon. Because the $\alpha'$ corrections reach their largest values on the horizon then having fixed the symmetries the H-boundary conditions guaranty that $\Psi_i^1(x)$ is bounded outside the horizon,
\begin{eqnarray}
\Psi_i^1(x)&<&\infty~,~ \forall  |x|\in [r_H,\infty)\,.
\end{eqnarray}
Second order linear differential equations have two solutions. In general the H-boundary conditions exclude one of them  and identify the other one. There exits no further freedom to impose more constraints on the solutions. Thus we conclude that:
\begin{quotation}
\hspace{-0.7cm} {\bf Lemma 2:} The H-boundary conditions do not necessarily retain the fall off of the fields at asymptotic infinity.
\end{quotation}
Note that these lemmas are not in contradiction with the supersymmetry. If we knew the $\alpha'$  corrections to the onshell supersymmetric constraints then we could have used the supersymmetric constraints rather the equations of motions  to obtain a set of non-homogeneous first order linear differential equations for the $\alpha'$ corrections to the background fields. Requiring the $\alpha'$ corrections to vanish at infinity fixes all the boundary conditions for these first order equations. Thus again we could have concluded that the fall off of the fields at asymptotic infinity might receive $\alpha'$ corrections. In addition we learn that the diverging modes on the horizon are non-supersymmetric fluctuations on the supersymmetric background.

For the Schwarzschild black hole, it is showed  \cite{martin,CallanRCMyers} that imposing the H-boundary conditions   produces corrections to the Newtonian mass of the black hole since the mass is given by the fall off of the time-time component of the canonical metric at asymptotic infinity. The fall off of the fields identifies the charges of the dyonic black hole. Thus the second lemma implies that  the charges of the dyonic black hole might get modified by the $\alpha'$ corrections. Also note that this does not necessarily contradicts the charge quantisation since the quanta of the charge could receive corrections. 

In the perturbative studies of the string scattering amplitudes one is allowed to redefine the fields,
\begin{eqnarray}\label{redefinition}
\tilde{\Psi}_i &=& \Psi_i\,+\, \alpha'\, \mathit{R_i}\,+\,O(\alpha'^2)
\end{eqnarray}
where $\mathit{R_i}$ are tensors of appropriate degree and dimension constructed from polynomials of $\Psi_i$ and their derivatives. The field redefinition alters  the induced action and subsequently  the equations of motion. For example a general field redefinition given by the eq.\,(\ref{redefinition}) changes the equations for the linear $\alpha'$ corrections (\ref{EQAlpha}) to,
\begin{eqnarray}\label{alpha-redefined}
\dfrac{\delta^2 S^0[\Psi]}{\delta \Psi_i\delta\Psi_j}\big{|}_{_{\alpha'=0}}\tilde{\Psi}^1_j &=&-\dfrac{\delta S^1[\Psi]}{\delta \Psi_i}\big{|}_{_{\alpha'=0}} - \dfrac{\delta^2 S^0[\Psi]}{\delta \Psi_i\delta\Psi_j}\big{|}_{_{\alpha'=0}}{\mathit{R_i}}\,. 
\end{eqnarray}
 The field redefinition ambiguity is related to the freedom in choosing different renormalisation schemes and regularisation methods  in the sigma model. Ref.\,\cite{Cvetic:1995yq} has considered the dyonic black as a generalisation of the null chiral sigma models \cite{Horowitz:1994rf}  and has proven that there exists a scheme in which $\Psi_i^0$ does not receive any $\alpha'$ corrections. This means that there exists $\mathit{R_i}=\mathit{R_i}^\star$ for which the right hand side of the eq.\,(\ref{alpha-redefined}) vanishes and $\tilde{\Psi}^1_j=0$ is the solution to the eq.\,(\ref{alpha-redefined}). Thus the solution in the scheme where the $\alpha'$ corrections are given by the eq.\,(\ref{EQAlpha}) reads 
\begin{eqnarray}
{\Psi}_i &=& \Psi_i^0\,-\, \alpha'\, \mathit{R_i}^\star\,+\,O(\alpha'^2)\,.
\end{eqnarray}
Since any field redefinition should contain two derivatives then the fall off of ${\Psi_i}$ at infinity, which is relevant for the charge identification, is the same as the one of $\Psi_i^0$ and we conclude that 
\begin{quotation}
\hspace{-0.7cm} {\bf Lemma 3:} There exists no $\alpha'$ correction to the charges of the dyonic black hole.
\end{quotation}

\section{The $\alpha'$ corrections to the entropy of dyons}
The near horizon configuration of the dyonic black hole at the supergravity approximation is $AdS_2\times S^2$. When the horizon is large the $\alpha'$ corrections do not change the geometry of the horizon. Therefore the near horizon configuration of the $\alpha'$ corrected dyonic black hole can be written in the following way
\begin{eqnarray}
ds^2 &=& \mathit{v_1} (-r^2 d\tau^2 + \frac{dr^2}{r^2}) + \mathit{v_2} (d\theta^2 + \sin^2 \theta d\phi^2)\,,\\
T_1 &=& \mathit{T_1} ~,~ T_2\,=\,\mathit{T_2}~,~e^{-2\Phi}\,=\,\mathit{s}\,,\\\label{guagefieldsstrengths}
F^1_{r\tau}&=& \mathit{e_1}~,~F^{3}_{r\tau}\,=\,\mathit{e_3}~,~F^2_{\theta\phi}\,=\,\frac{\mathit{p_2}}{4\,\mathit{\pi}}\sin\theta
~,~F^4_{\theta\phi}\,=\,\frac{\mathit{p_4}}{4\,\mathit{\pi}}\sin\theta\,,
\end{eqnarray}
where the horizon in located at $r=0$ and $\mathit{v_1},\cdots,\mathit{p_4}$ are constant parameters labelling the horizon. Note that $\mathit{v_1}$ and $\mathit{v_2}$ are constant due to the geometry of the horizon and $\mathit{T_1}$, $\mathit{T_2}$, $\mathit{s}$ are constant since they represent the limit of $r\to 0$ of the scalars. $\mathit{e_1}$, $\mathit{e_3}$, $\mathit{p_2}$ and $\mathit{p_4}$ are constant due to the coordinates chosen to represent the background and in accordance with the supergravity approximation.

We have shown  that the part of the induced action which we are intersted in is a functional of the gauge field strengths but not of the gauge fields themselves. Thus the entropy formalism techniques \cite{int4,int5} can be employed to express the parameters of the near horizon configuration in terms of the charges of the dyonic black hole. The entropy formalism utilises the entropy function defined by 
\begin{eqnarray}
f(\vec{\mathit{v}},\vec{\mathit{T}},\vec{\mathit{e}},\vec{\mathit{p}},\mathit{s})&=&\frac{1}{32\,\pi}\, \int d\theta d\phi \sqrt{-\det g}\, \mathit{s}\, L(\vec{\mathit{v}},\vec{\mathit{T}},\vec{\mathit{e}},\vec{\mathit{p}})\,, 
\end{eqnarray}
where $L(\vec{\mathit{v}},\vec{\mathit{T}},\vec{\mathit{e}},\vec{\mathit{p}})$ is the induced Lagrangian (\ref{inducedlagrangian}) evaluated on the horizon configuration. The equations of motions are equivalent to 
\begin{eqnarray}\label{theEQ}
\frac{\partial f(\vec{\mathit{v}},\vec{\mathit{T}},\vec{\mathit{e}},\vec{\mathit{p}},\mathit{s})}{\partial \mathit{v_i}} &=&0\,,\\
\nonumber
\frac{\partial f(\vec{\mathit{v}},\vec{\mathit{T}},\vec{\mathit{e}},\vec{\mathit{p}},\mathit{s})}{\partial \mathit{T_i}} &=&0\,,\\
\nonumber
\frac{\partial f(\vec{\mathit{v}},\vec{\mathit{T}},\vec{\mathit{e}},\vec{\mathit{p}},\mathit{s})}{\partial \mathit{s}} &=&0\,,\\
\nonumber
\frac{\partial f(\vec{v},\vec{\mathit{T}},\vec{e},\vec{p},\mathit{s})}{\partial \mathit{e_i}} &=& \mathit{q_{i}} ~,~i\in\{1,3\}\nonumber
\end{eqnarray}
where $\mathit{q_1}$ and $\mathit{q_3}$ are the electric charges of the dyonic black hole and $\mathit{p_2}$ and $\mathit{p_4}$ are identified as the magnetic charges. In the following we first obtain these equations and next we find their $\alpha'$ perturbative solutions.

To evaluate the induced action on the horizon configuration we use the eq. (\ref{themetric}), (\ref{theB}) and (\ref{thedilaton}) to reconstruct the ten dimensional fields for the near horizon configuration,
\begin{eqnarray}\label{tendg}
d\boldsymbol{s}^2&=& ds^2 + \mathit{T_1} (dy_1\,+\,2\, \mathit{e_1}\, r\, dt)^2 + \mathit{T_2} (dy_2\,-\,\frac{\mathit{p_2}}{2\pi}\,\cos\theta\, d\phi)^2\,+\,dz_m^2\,,\\\label{tendB}
\boldsymbol{B} &=& 2\,\mathit{e_3}\, r\, dt\wedge dy_1\, -\,\frac{\mathit{p_4}}{2\pi}\,\cos\theta\, d\phi\wedge dy_2\,,\\
e^{-2\boldsymbol{\phi}} &=&  {\displaystyle\frac{\mathit{s}} {\mathit{V}\,\mathit{T_1}\, \mathit{T_2}}}\,,
\end{eqnarray}
where $\mathit{V}$ is the volume of the compactified space and the gauges  (\ref{guagefieldsstrengths}) are chosen by
\begin{eqnarray}
A^1_{\mu} &=& [\mathit{e_1}\, r, 0, 0, 0]~,~ A^3_{\mu} \,=\, [\mathit{e_3}\, r, 0,0,0]\,,\\
A^2_{\mu} &=& [0, 0, 0, -\frac{\mathit{p_2}}{4\pi} \cos\theta]~,~ A^4_{\mu} \,=\, [0, 0, 0, -\frac{\mathit{p_4}}{4\pi} \cos\theta]\,.
\end{eqnarray}
In the ten dimensions using the ten dimensional fields - the eq. (\ref{tendg}) and (\ref{tendB})- one finds that
\begin{eqnarray}
\mathit{L_0}&=&{\boldsymbol{R}_{\scriptscriptstyle{\text{Ricci}}}} - \frac{1}{12}\boldsymbol{H}_{ijk}\boldsymbol{H}^{ijk}\,=\,\\ &=&
- {\displaystyle \frac {2}{\mathit{v_1}}} 
+ {\displaystyle \frac {2}{\mathit{v_2}}}
+ {\displaystyle \frac {2\mathit{e_1}^{2}\mathit{T_1}^{2}}{\mathit{v_1}^{2}}}
+ {\displaystyle \frac {2\mathit{e_3}^{2}}{\mathit{v_1}^{2}\mathit{T_1}^{2}}}  
-{\displaystyle \frac {\mathit{p_2}^{2}\mathit{T_2}^{2}}{8\mathit{v_2}^{2}\pi ^{2}}} 
- {\displaystyle \frac {\mathit{p_4}^{2}}{8\mathit{v_2}^{2}\pi ^{2}\mathit{T_2}^{2}}},
\nonumber
\\
\nonumber
\\
\mathit{L_1}&=&
\frac{1}{8}{\boldsymbol{R}_{klmn}\boldsymbol{R}^{klmn}} \,=\,\\
&=&
+ {\displaystyle \frac {1}{2\,\mathit{v_1}^{2}}} 
+ {\displaystyle \frac {1}{2\,\mathit{v_2}^{2}}} 
- {\displaystyle \frac {3\mathit{e_1}^{2}\mathit{T_1}^{2}}{\mathit{v_1}^{3}}}
- {\displaystyle\frac {3\mathit{p_2}^{2}\mathit{T_2}^{2}}{16\,\mathit{v_2}^{3}\pi ^{2}}} 
+ {\displaystyle \frac {11\mathit{T_1}^{4}\mathit{e_1}^{4}}{2\,\mathit{v_1}^{4}}} 
+ {\displaystyle \frac {11\mathit{p_2}^{4}\mathit{T_2}^{4}}{512\,\mathit{v_2}^{4}\pi ^{4}}}  
\,, 
\nonumber\\
\nonumber\\
\mathit{L_2}&=&-\frac{1}{16}
\boldsymbol{R}_{klmn}\boldsymbol{H}_{p}^{~kl}\boldsymbol{H}^{pmn} \,=\,\\ \nonumber
&=&
- {\displaystyle \frac {\mathit{e_3}^{2}}{\mathit{v_1}^{3}\,\mathit{T_1}^{2}}} 
- {\displaystyle \frac {\mathit{p_4}^{2}}{16\,\pi ^{2}\,\mathit{v_2}^{3}\,\mathit{T_2}^{2}}}
+ {\displaystyle \frac {\mathit{e_3}^{2}\,\mathit{e_1}^{2}}{\mathit{v_1}^{4}}}  
+ {\displaystyle \frac {\mathit{p_4}^{2}\,\mathit{p_2}^{2}}{256\,\pi ^{4}\,\mathit{v_2}^{4}}}\,
, 
\\
\mathit{L_3}&=&-\frac{1}{64}
\boldsymbol{H}_{k}^{~mn}\boldsymbol{H}_{lmn}\boldsymbol{H}^{kpq}\boldsymbol{H}^{l}_{~pq}\,=\,
- {\displaystyle \frac {3\,\mathit{e_3}^{4}}{\mathit{v_1}^{4}\,\mathit{T_1}^{4}}}  
- {\displaystyle \frac{3\,\mathit{p_4}^{4}}{256\,\pi^{4}\,\mathit{v_2}^{4}\,\mathit{T_2}^{4}}}\,, \\
\mathit{L_4}&=&
\frac{1}{192}\boldsymbol{H}_{klm}\boldsymbol{H}^{k}_{~pq}\boldsymbol{H}_{r}^{~lp}\boldsymbol{H}^{rmq}\,=\,
{\displaystyle \frac {\mathit{e_3}^{4}}{2\,\mathit{v_1}^{4}\,\mathit{T_1}^{4}}}
 + {\displaystyle \frac {\mathit{p_4}^{4}}{512\,\pi ^{4}\,\mathit{v_2}^{4}\,\mathit{T_2}^{4}}} \,.
\end{eqnarray}
Inserting the above expressions in the ten dimensional action at the supergravity approximation (\ref{treelevelaction}) and its linear $\alpha'$ corrections (\ref{LMT}) gives
\begin{eqnarray}
S&=& \boldsymbol{S}\,=\,\frac{1}{32\pi}\int\,dt\,dr\,d\phi\,d\cos\theta~ \mathit{s}\,\mathit{v_1}\,\mathit{v_2}\,(\mathit{L_0}\,+\,\alpha'(\mathit{L_1}+\mathit{L_2}+\mathit{L_3}+\mathit{L_4})) + O(\alpha'^2)\,,\nonumber\\ 
\end{eqnarray}
where the integration over the compactified space is understood.\footnote{The idea of identifying the induced action near the horizon rather than in the whole of the space was used in the ref.\,\cite{Exirifard:2006wa} to study  the linear $\alpha'$ corrections for singular backgrounds representing a wrapped fundamental string.} Then the entropy function reads
\begin{eqnarray}\label{falpha}
f(\vec{\mathit{v}},\vec{\mathit{T}},\vec{\mathit{e}},\vec{\mathit{p}},\mathit{s}) & = &\frac{1}{8}\mathit{s}\,\mathit{v_1}\,\mathit{v_2}\,(\mathit{L_0}\,+\,\alpha'(\mathit{L_1}+\mathit{L_2}+\mathit{L_3}+\mathit{L_4})) + O(\alpha'^2)\,,  
\end{eqnarray}
using which in the eq.\,(\ref{theEQ}) gives the equations of motion near the horizon. These equations should be solved perturbatively. Thus we write $\alpha'$ expansion series for the constant parameters labelling the horizon configuration,
\begin{eqnarray}\label{AH}
\mathit{v_i}&=& \mathit{v_i}^0\, (1\,+\, \alpha'\,\tilde{\mathit{v_i}}\, +\, O(\alpha'^2))\,,\\
\nonumber
\mathit{T_i}&=& \mathit{T_i}^0\, (1\, + \alpha'\,\tilde{\mathit{T_i}}\, +\, O(\alpha'^2))\,,\\
\nonumber
\mathit{e_i}&=& \mathit{e_i}^0\, (1\, + \alpha'\,\tilde{\mathit{e_i}}\, +\, O(\alpha'^2))\,,\\
\nonumber
\mathit{s}&=& \mathit{s}^0\, (1\, + \alpha'\,\tilde{\mathit{s}}\, +\, O(\alpha'^2))\,.
\end{eqnarray}
Not that the electric and magnetic charges are fixed to their values at the supergravity approximation because there exists no $\alpha'$ corrections to the charges of the dyonic black hole.
The equations of motion (\ref{theEQ}) at the supergravity approximation ($\alpha'=0$) are solved by
\begin{eqnarray}\label{HSA}
\mathit{v_1}^0 &=& \mathit{v_2}^0\,=\, \frac{\mathit{p_2}\,\mathit{p_4}}{4\,\pi^2}\,,\\ 
\nonumber
\mathit{T_1}^0 &=& \sqrt{\frac{\mathit{p_4}}{\mathit{p_2}}}~,~
\nonumber
\mathit{T_2}^0 \,=\,\sqrt{\frac{\mathit{q_1}}{\mathit{q_3}}}\,,\\
\nonumber
\mathit{e_1}^0 &=& \frac{1}{4\pi}\,\sqrt{\frac{{\mathit{q_3}\,\mathit{p_2}\,\mathit{p_4}}}{\mathit{q_1}}}~,~
\nonumber
\mathit{e_3}^0 \,=\, \frac{1}{4\pi}\,\sqrt{\frac{{\mathit{q_1}\,\mathit{p_2}\,\mathit{p_4}}}{\mathit{q_3}}}\,,\\
\nonumber
\mathit{s}^0 &=& 8 \,\pi \, \sqrt{\frac{\mathit{q_1}\,\mathit{q_3}}{\mathit{p_2}\,\mathit{p_4}}}\,.
\end{eqnarray}
These are the horizon configuration parameters at the supergravity approximation. Inserting the eq. (\ref{AH}) and (\ref{HSA}) in the eq.\,(\ref{theEQ}) gives a set of linear algebraic equations for the linear $\alpha'$ corrections (\ref{AH}) to the supergravity approximation (\ref{HSA}). These linear equations are solved by 
\begin{eqnarray}
\tilde{\mathit{v}}_{\mathit{1}} &=& 0\,,\\
\frac{1}{2}\tilde{\mathit{v}}_{\mathit{2}}&=&\tilde{\mathit{T}}_{\mathit{2}}\,=\,-\,\tilde{\mathit{T}}_{\mathit{1}}\,=\,-\,\tilde{\mathit{s}}\,=\,
\tilde{\mathit{e}}_{\mathit{1}} \,=\,\tilde{\mathit{e}}_{\mathit{3}}\,=\,\frac{\pi^2}{\mathit{p_2}\,\mathit{p_4}}\,.
\end{eqnarray}
The modified Hawking-Bekenstein (Wald) entropy is expressed by the Legendre transformation of the entropy function
\begin{eqnarray}\label{Waldentropy}
S_{bh} &=& 2\,\pi\,(\mathit{e_1}\,\frac{\partial f(\vec{\mathit{v}},\vec{\mathit{T}},\vec{\mathit{e}},\vec{\mathit{p}},\mathit{s})}{\partial\mathit{e_1}}+\mathit{e_3}\,\frac{\partial f(\vec{\mathit{v}},\vec{\mathit{T}},\vec{\mathit{e}},\vec{\mathit{p}},\mathit{s})}{\partial\mathit{e_3}}-f(\vec{\mathit{v}},\vec{\mathit{T}},\vec{\mathit{e}},\vec{\mathit{p}},\mathit{s}))\,, 
\end{eqnarray}
evaluated on the horizon configuration \cite{int4,int5}. Inserting the eq. (\ref{AH})  and  (\ref{HSA})  in the eq.\,(\ref{Waldentropy}) we get 
\begin{eqnarray}\label{WaldEntropy1}
S_{bh} &=& \sqrt{\mathit{p_2}\,\mathit{p_4}\,\mathit{q_1}\,\mathit{q_3}}\,(1\,+\,\frac{\pi^2\,\alpha'}{\mathit{p_2}\,\mathit{p_4}})\,+\,O(\alpha'^2)\,. 
\end{eqnarray}
We notice that Wald entropy does not depend on the values of $\tilde{\mathit{v}}_{\mathit{1}}$, $\tilde{\mathit{v}}_{\mathit{2}}$, $\tilde{\mathit{T}}_{\mathit{1}}$, $\tilde{\mathit{T}}_{\mathit{2}}$, $\tilde{\mathit{e}}_{\mathit{1}}$, $\tilde{\mathit{e}}_{\mathit{2}}$ and $\tilde{\mathit{s}}$.

To find all the linear $\alpha'$ corrections to the entropy we should add the contribution of the gravitational Chern-Simons term to the eq.\,(\ref{WaldEntropy1}). For the supersymmetric dyon the gravitational Chern-Simons corrections to the entropy reads \cite{Bindusar}
\begin{eqnarray}
S^{(1)}_{CS} &=& \sqrt{\mathit{p_2}\,\mathit{p_4}\,\mathit{q_1}\,\mathit{q_3}}\,(\frac{\pi^2\,\alpha'}{\mathit{p_2}\,\mathit{p_4}})\,+\,O(\alpha'^2)\,. 
\end{eqnarray}
Adding this to the eq.\,(\ref{WaldEntropy1}) we find 
\begin{eqnarray}\label{WaldEntropy}
S_{BH} &=& \sqrt{\mathit{p_2}\,\mathit{p_4}\,\mathit{q_1}\,\mathit{q_3}}\,(1\,+\,2\,\frac{\pi^2\,\alpha'}{\mathit{p_2}\,\mathit{p_4}})\,+\,O(\alpha'^2)\,, 
\end{eqnarray}
as all the linear $\alpha'$ corrections to the entropy.

In the following we would like to compare all the linear  $\alpha'$ corrections to the entropy (\ref{WaldEntropy}) with the corrections given by only the inclusion of the Gauss-Bonnet Lagrangian in the induced action \cite{int4}. The square of the Riemann tensor of the four dimensional metric is 
\begin{eqnarray}\label{R2}
\frac{1}{8}\,R_{ijkl}\,R^{ijkl} & = & \frac{1}{2\,\mathit{v_1}^2} \,+\, \frac{1}{2\,\mathit{v_2}^2}\,. 
\end{eqnarray}
We see that the eq.\,(\ref{R2}) coincides with the first two terms in $\mathit{L_1}$. In total ten terms in $\mathit{L_1}+\mathit{L_2}+\mathit{L_3}+\mathit{L_4}$ are not given by the square of the Riemann tensor. The Gauss-Bonnet  Lagrangian in the four dimensions is
\begin{eqnarray}
L_{GB} &=& R_{ijkl}\,R^{ijkl} \,-\,4\,R_{ij}\,R^{ij}\,+\,R^2\,=\,-\,\frac{8}{\mathit{v_1}\,\mathit{v_2}}\,.
\end{eqnarray}
Including the Gauss-Bonnet action to the induced action at the supergravity approximation is equal to including the eq.\,(\ref{R2}) and performing a field redefinition. Thus ten terms in the linear $\alpha'$ corrections to the induced action are not produced by the inclusion of the Gauss-Bonnet action. The inclusion of the Gauss-Bonnet action in the induced action at supergravity approximation gives the following entropy function  
\begin{eqnarray}\label{fGB}
f^\star &=& \frac{1}{8}\, \mathit{s}^\star\,\mathit{v_1}^\star\,\mathit{v_2}^\star\,(\mathit{L_0}\,+\,\frac{\alpha'}{8}\,L_{GB})
\,=\,\frac{1}{8}\,  \mathit{s}^\star\,\mathit{v_1}^\star\,\mathit{v_2}^\star\,\mathit{L_0}\,-\,\frac{\alpha'}{8}\,\mathit{s}^\star \,+\,O(\alpha')\,,
\end{eqnarray}
where we used the superscript $^\star$ to distinguish the near horizon parameters identified by the eq.\,(\ref{fGB}) with those identified by the eq.\,(\ref{falpha}).  This entropy function (\ref{fGB}) identifies the horizon configuration parameters to
\begin{eqnarray}
 \frac{\mathit{v_i}^\star}{\mathit{v_i}^0}\,=\,\frac{\mathit{e_i}^\star}{\mathit{e_i}^0}\,=\,1\,+\frac{2\,\pi^2 \,\alpha'}{\mathit{p_2}\,\mathit{p_4}}\,,~\frac{\mathit{T_i}^\star}{\mathit{T_i}^0}=1\,,~\frac{\mathit{s}^\star}{\mathit{s}^0}\,=\,1\,-\frac{2\pi^2 \,\alpha'}{\mathit{p_2}\,\mathit{p_4}}
\end{eqnarray}
where $v_1^0$, $\cdots$, $s^0$ are given by eq.\,(\ref{HSA}) and for which the entropy reads
\begin{eqnarray}\label{StatEntropy}
{S}_{GB} &=& \sqrt{\mathit{p_2}\,\mathit{p_4}\,\mathit{q_1}\,\mathit{q_3}}\,(1\,+\,\frac{2\,\pi^2\,\alpha'}{\mathit{p_2}\,\mathit{p_4}})\,+\,O(\alpha')\,, 
\end{eqnarray}
Ref.\,\cite{int4} has included the Gauss-Bonnet action in the induced action at the supergravity approximation and has solved the corresponding truncated $\alpha'$-corrected  equations of motion exactly. We note that the eq.\,(\ref{StatEntropy}) is in agreement with the large charge expansion of eq.\,(3.13) of the ref.\,\cite{int4} after setting $n=2\,\mathit{q_1}$, $w=2\,\mathit{q_3}$, $\tilde{N}=\frac{\mathit{p_2}}{4\,\pi}$, $\tilde{W}=\frac{\mathit{p_4}}{4\,\pi}$ and using the unit of $\alpha'=16$.

 We notice that the eq.\,(\ref{StatEntropy}) is  in agreement with the eq.\,(\ref{WaldEntropy}) therefore we verify that all the linear $\alpha'$ corrections to the entropy of a supersymetric dyon with a large horizon can be reproduced by  the inclusion of only the Gauss-Bonnet Lagrangian to its Lagrangian density at the supergravity approximation. This is in accordance with the non-renormalisable theorems of the ref.\,\cite{Kraus1,Kraus2} and the conclusion of the ref.\,\cite{Bindusar}.

Also note that the inclusion of only the Gauss-Bonnet Lagrangian to the supergravity approximation is in agreement \cite{int4}  with  the inclusion of the supersymmetric version of the square of the Riemann tensor \cite{Behrndt:1998eq}-\cite{LopesCardoso:2000fp} which  in turn is in agreement with the large charge expansion series of the statistical entropy \cite{Dijkgraaf:1996it}-\cite{David:2006yn}, and thus all the linear $\alpha'$ corrections to the entropy are in agreement with the statistical entropy.

\section*{Acknowledgements}
I would like to thank Loriano Bonora and Martin O'loughlin for discussions throughout the work and careful reading of the draft. I would like to thank Justin David for discussions. I thank Marco Serone for discussions on the second section.


\begin{thebibliography}{10}

\bibitem{Bindusar}
B. Sahoo and A. Sen, {\it $\alpha'$ corrections to extremal dyonic black holes in heterotic string theories},
\href{http://xxx.lanl.gov/abs/hep-th/0608182}{{\tt hep-th/0608182}}.

\bibitem{Dijkgraaf:1996it}
R.~Dijkgraaf, E.~P. Verlinde, and H.~L. Verlinde, {\it Counting dyons in
  \protect{N = 4} string theory},  {\em Nucl. Phys.} {\bf B484} (1997)
  543--561, [\href{http://xxx.lanl.gov/abs/hep-th/9607026}{{\tt
  hep-th/9607026}}].

\bibitem{LopesCardoso:2004xf}
G.~Lopes~Cardoso, B.~de~Wit, J.~Kappeli, and T.~Mohaupt, {\it Asymptotic
  degeneracy of dyonic \protect{N = 4} string states and black hole entropy},
  {\em JHEP} {\bf 12} (2004) 075,
  [\href{http://xxx.lanl.gov/abs/hep-th/0412287}{{\tt hep-th/0412287}}].

\bibitem{Shih:2005uc}
D.~Shih, A.~Strominger, and X.~Yin, {\it Recounting dyons in \protect{N = 4}
  string theory}, {\em JHEP} {\bf 0610} (2006) 087, [\href{http://xxx.lanl.gov/abs/hep-th/0505094}{{\tt
  hep-th/0505094}}].

\bibitem{Gaiotto:2005hc}
D.~Gaiotto, {\it Re-recounting dyons in \protect{N = 4} string theory},
  \href{http://xxx.lanl.gov/abs/hep-th/0506249}{{\tt hep-th/0506249}}.

\bibitem{Shih:2005he}
D.~Shih and X.~Yin, {\it Exact black hole degeneracies and the topological
  string},  {\em JHEP} {\bf 04} (2006) 034,
  [\href{http://xxx.lanl.gov/abs/hep-th/0508174}{{\tt hep-th/0508174}}].

\bibitem{David:2006yn}
J.~R. David and A.~Sen, {\it \protect{CHL} dyons and statistical entropy
  function from \protect{D1-D5} system}, {\em JHEP} {\bf 0611} (2006) 072
  [\href{http://xxx.lanl.gov/abs/hep-th/0605210}{{\tt hep-th/0605210}}].

\bibitem{Behrndt:1998eq}
K.~Behrndt {\em et~al.}, {\it Higher-order black-hole solutions in \protect{N =
  2} supergravity and \protect{Calabi-Yau} string backgrounds},  {\em Phys.
  Lett.} {\bf B429} (1998) 289--296,
  [\href{http://xxx.lanl.gov/abs/hep-th/9801081}{{\tt hep-th/9801081}}].

\bibitem{22}
G.~L. Cardoso, B.~de~Wit, and T.~Mohaupt, {\it Corrections to macroscopic
  supersymmetric black-hole entropy},  {\em Phys.Lett.} {\bf B451} (1999) 309,
  [\href{http://xxx.lanl.gov/abs/hep-th/9812082}{{\tt hep-th/9812082}}].

\bibitem{23}
G.~L. Cardoso, B.~de~Wit, and T.~Mohaupt, {\it Deviations from the area law for
  supersymmetric black holes},  {\em Fortsch.Phys.} {\bf 48} (2000) 49,
  [\href{http://xxx.lanl.gov/abs/hep-th/9904005}{{\tt hep-th/9904005}}].

\bibitem{LopesCardoso:1999ur}
G.~Lopes~Cardoso, B.~de~Wit, and T.~Mohaupt, {\it Macroscopic entropy formulae
  and non-holomorphic corrections for supersymmetric black holes},  {\em Nucl.
  Phys.} {\bf B567} (2000) 87--110,
  [\href{http://xxx.lanl.gov/abs/hep-th/9906094}{{\tt hep-th/9906094}}].

\bibitem{25}
G.~L. Cardoso, B.~de~Wit, and T.~Mohaupt, {\it Area law corrections from state
  counting and supergravity},  {\em Class.Quant.Grav.} {\bf 17} (2000) 1007,
  [\href{http://xxx.lanl.gov/abs/hep-th/9910179}{{\tt hep-th/9910179}}].

\bibitem{26}
T.~Mohaupt, {\it Black hole entropy, special geometry and strings},  {\em
  Fortsch.Phys.} {\bf 49} (2001) 3,
  [\href{http://xxx.lanl.gov/abs/hep-th/0007195}{{\tt hep-th/0007195}}].

\bibitem{LopesCardoso:2000qm}
G.~Lopes~Cardoso, B.~de~Wit, J.~Kappeli, and T.~Mohaupt, {\it Stationary
  \protect{BPS} solutions in \protect{N = 2} supergravity with
  \protect{$R^2$}-interactions},  {\em JHEP} {\bf 12} (2000) 019,
  [\href{http://xxx.lanl.gov/abs/hep-th/0009234}{{\tt hep-th/0009234}}].

\bibitem{LopesCardoso:2000fp}
G.~Lopes~Cardoso, B.~de~Wit, J.~Kappeli, and T.~Mohaupt, {\it Examples of
  stationary \protect{BPS} solutions in \protect{N = 2} supergravity theories
  with \protect{$R^2$}-interactions},  {\em Fortsch. Phys.} {\bf 49} (2001)
  557--563, [\href{http://xxx.lanl.gov/abs/hep-th/0012232}{{\tt
  hep-th/0012232}}].

\bibitem{wald}
R.~M. Wald, {\it Black hole entropy is \protect{Noether} charge},  {\em
  Phys.Rev.} {\bf D48} (1993) 3427,
  [\href{http://xxx.lanl.gov/abs/gr-qc/9307038}{{\tt gr-qc/9307038}}].

\bibitem{wald2}
V.~Iyer and R.~M. Wald, {\it Some properties of \protect{Noether} charge and a
  proposal for dynamical black hole entropy},  {\em Phys.Rev.} {\bf D50} (1994)
  846, [\href{http://xxx.lanl.gov/abs/gr-qc/9403028}{{\tt gr-qc/9403028}}].

\bibitem{int4}
A.~Sen, {\it Entropy function for \protect{Heterotic} black holes},
  {\em JHEP} {\bf 0603}, 008 (2006), [\href{http://xxx.lanl.gov/abs/hep-th/0508042}{{\tt hep-th/0508042}}].


\bibitem{Kraus1}
P. Kraus and F. Larsen, {\it Microscopic black hole entropy in theories with higher derivatives}, {\em JHEP} {\bf 0509}, 034 (2005),[\href{http://xxx.lanl.gov/abs/hep-th/0506176}{{\tt hep-th/0506176}}].


\bibitem{Kraus2}
P. Kraus and F. Larsen, {\it Holographic gravitational anomalies}, {\em JHEP} {\bf 0601}, 022 (2006),[\href{http://xxx.lanl.gov/abs/hep-th/0508218}{{\tt hep-th/0508218}}].

\bibitem{Qasem}
Qasem Exirifard, {\it The world-sheet corrections to dyons in the Heterotic theory},
[\href{http://arxiv.org/abs/hep-th/0607094v2}{{\tt hep-th/0607094v1}}].

\bibitem{Bindusar2}
B. Sahoo and A. Sen, {\it BTZ black hole with Chern-Simons and higher derivative terms},
\href{http://xxx.lanl.gov/abs/hep-th/0601228}{{\tt hep-th/0601228}}.


\bibitem{Cvetic:1995uj}
M.~Cvetic and D.~Youm, {\it Dyonic bps saturated black holes of heterotic
  string on a six torus},  {\em Phys. Rev.} {\bf D53} (1996) 584--588,
  [\href{http://xxx.lanl.gov/abs/hep-th/9507090}{{\tt hep-th/9507090}}].

\bibitem{stringamplitude1}
C.~Hull and P.~Townsend, {\it The two loop $\beta$ function for $\sigma$ model
  with torsion},  {\em Phys. Lett.} {\bf B191} (1987) 115.

\bibitem{stringamplitude2}
R.~Metsaev and A.~Tseytlin, {\it Order $\alpha'$ (two-loop) equivalence of the
  string equations of motion and the $\sigma$ model weyl invarince conditions},
   {\em Nucl.Phys.} {\bf B293} (1987) 385.

\bibitem{Cvetic:1995yq}
M.~Cvetic and A.~A. Tseytlin, {\it General class of \protect{BPS} saturated
  dyonic black holes as exact superstring solutions},  {\em Phys. Lett.} {\bf
  B366} (1996) 95--103, [\href{http://xxx.lanl.gov/abs/hep-th/9510097}{{\tt
  hep-th/9510097}}].

\bibitem{Horowitz:1994rf}
G.~T. Horowitz and A.~A. Tseytlin, {\it A new class of exact solutions in
  string theory},  {\em Phys. Rev.} {\bf D51} (1995) 2896--2917,
  [\href{http://xxx.lanl.gov/abs/hep-th/9409021}{{\tt hep-th/9409021}}].

\bibitem{Ferrara:1995ih}
S.~Ferrara, R.~Kallosh, and A.~Strominger, {\it \protect{N = 2} extremal black
  holes},  {\em Phys. Rev.} {\bf D52} (1995) 5412--5416,
  [\href{http://xxx.lanl.gov/abs/hep-th/9508072}{{\tt hep-th/9508072}}].

\bibitem{Strominger:1996kf}
A.~Strominger, {\it Macroscopic entropy of \protect{N=2} extremal black holes},  {\em
  Phys. Lett.} {\bf B383} (1996) 39--43,
  [\href{http://xxx.lanl.gov/abs/hep-th/9602111}{{\tt hep-th/9602111}}].

\bibitem{int5}
A.~Sen, {\it Black hole entropy function and the attractor mechanism in higher
  derivative gravity},  {\em JHEP} {\bf 09} (2005) 038,
  [\href{http://xxx.lanl.gov/abs/hep-th/0506177}{{\tt hep-th/0506177}}].

\bibitem{5Gross}
D. J. Gross, J. A. Harvey, E. Martinec, and R. Rohm, {\it Heterotic string theory: (ii). The interacting heterotic string theory}, {\em Nucl. Phys.} {\bf B198} (1982) 474.

\bibitem{5Nishino}
D. Chang and H. Nishino, {\it Heterotic-string O($\alpha'$) corrections to D=10, N=1 supergravity}, {\em Phys. Lett.} {\bf 179B} (1986) 75.

\bibitem{5Cai}
Y. Cai and C. Nunez, {\it Heterotic string covariant amplitudes and low-energy effective action},{\em Nucl. Phys.} {\bf B287} (1987) 279.

\bibitem{5Green}
M. B. Green and J. H. Schwarz, {\it Supersymmetric dual string theory: (ii). Vertices and trees},{\em Nucl. Phys.} {\bf B198} (1986) 75.

\bibitem{5Schwarz}
J. H. Schwarz, {\it Superstring theory}, {\em Phys. Rep.} {\bf 89} (223) 1982.

\bibitem{vertex}
S. Weinberge, {\it Coupling constants and vertex functions in string theories}, {\em Phys. Lett.} {\bf 156B} (1985) 309.

\bibitem{DBranes}
M. B. Green and J. H. Schwarz, {\it Anomaly cancellation in supersymmetric $d=10$ gauge theory and superstring theory}, {\em Phys. Lett.} {\bf B149}(1984) 117.

\bibitem{SenOdd}
A.~Sen, {\it Strong-weak coupling duality in four dimensional string theory},
  {\em Int. J. Mod. Phys.} {\bf A 9} (1994) 3707,
  [\href{http://xxx.lanl.gov/abs/hep-th/9402002}{{\tt hep-th/9402002}}].

\bibitem{martin}
Q.~Exirifard and M.~O'Loughlin, {\it Two and three loop $\alpha'$ corrections
  to \protect{T}-duality: \protect{Kasner} and \protect{Schwarzschild}},  {\em
  JHEP} {\bf 12} (2004) 023,
  [\href{http://xxx.lanl.gov/abs/hep-th/0408200}{{\tt hep-th/0408200}}].

\bibitem{exir2}
Qasem~Exirifard, {\it Quadratic $\alpha'$ corrections to \protect{T}-duality},
  {\em JHEP} {\bf 07} (2005) 047,
  [\href{http://xxx.lanl.gov/abs/hep-th/0504133}{{\tt hep-th/0504133}}].

\bibitem{T1}
P.~E. Haagensen and K.~Olsen, {\it \protect{T}-duality and two-loop
  renormalization flows},  {\em Nucl.Phys.} {\bf B504} (1997) 326,
  [\href{http://xxx.lanl.gov/abs/hep-th/9704157.}{{\tt hep-th/9704157.}}].

\bibitem{Seif}
S.~Randjbar-Daemi and G.~Thompson, {\it Lecture notes on quantum field theory}
  .

\bibitem{CallanRCMyers}
C.~G. Callan, C.~Myers, R., and J.~Perry, M., {\it Black holes in string
  theory},  {\em Nucl. Phys.} {\bf B311} (1988) 673.

\bibitem{Exirifard:2006wa}
Q.~Exirifard, {\it The $\alpha'$ stretched horizon in the heterotic string}, {\em JHEP} {\bf 0610} (2006) 070
  [\href{http://xxx.lanl.gov/abs/hep-th/0604021}{{\tt hep-th/0604021}}].

\bibitem{Jacobson:1993vj}
T.~Jacobson, G.~Kang, and R.~C. Myers, {\it On black hole entropy},  {\em Phys.
  Rev.} {\bf D49} (1994) 6587,
  [\href{http://xxx.lanl.gov/abs/gr-qc/9312023}{{\tt gr-qc/9312023}}].

\bibitem{peet}
A.~W. Peet, {\it Entropy and supersymmetry of \protect{D}-dimensional extremal
  electric black holes versus string states},  {\em Nucl.Phys.} {\bf B456}
  (1995) 732, [\href{http://xxx.lanl.gov/abs/hep-th/9506200}{{\tt
  hep-th/9506200}}].

\bibitem{atish}
A.~Dabholkar, {\it Exact counting of black hole microstates},  {\em
  Phys.Rev.Lett.} {\bf 94} (2005) 241301,
  [\href{http://xxx.lanl.gov/abs/hep-th/0409148}{{\tt hep-th/0409148}}].





\end{thebibliography}
\providecommand{\href}[2]{#2}\begingroup\raggedright

\endgroup

\end{document}